# Spontaneous Formation of a Superconductor-Topological Insulator-Normal Metal Layered Heterostructure

By *Yu-Qi Wang, Xu Wu, Ye-Liang Wang\*, Yan Shao, Tao Lei, Jia-Ou Wang, Shi-Yu Zhu, Haiming Guo, Ling-Xiao Zhao, Gen-Fu Chen, Simin Nie, Hong-Ming Weng, Kurash Ibrahim, Xi Dai, Zhong Fang, Hong-Jun Gao\**

Prof. Hong-Jun Gao, Prof. Yeliang Wang, Prof. *Zhong Fang*, Prof. *Xi Dai*, Prof. *Gen-Fu Chen*, Dr. *Haiming Guo,* Dr. *Hong-Ming Weng, Yu-Qi Wang, Xu Wu, Yan Shao, Shi-Yu Zhu, Ling-Xiao Zhao, Simin Nie,*

Beijing National Laboratory of Condensed Matter Physics, Institute of Physics,

Chinese Academy of Sciences, Beijing 100190, P. R. China

E-mails: hjgao@iphy.ac.cn, ylwang@iphy.ac.cn

Prof. *Kurash Ibrahim*,  Dr. *Jia-Ou Wang, Tao Lei,*

Institute of High Energy Physics, Chinese Academy of Sciences, Beijing 100049, P. R. China.

Prof. Hong-Jun Gao, Prof. Yeliang Wang, Prof. *Zhong Fang*, Prof. *Xi Dai*, Prof. *Gen-Fu Chen,* Dr. *Hong-Ming Weng,*

Collaborative Innovation Center of Quantum Matter, Beijing 100084, China.

Y.Q.W. and X.W. contributed equally to this work.

The manuscript was written through contributions of all authors.

The discovery of graphene has spurred vigorous investigation of 2D materials, revealing a wide range of extraordinary properties and functionalities. 2D heterostructural materials have recently been fabricated by assembling isolated planes layer-by-layer in a desired sequence. Unusual properties and novel physical phenomena have been unveiled in such layered heterostructures. [1-4] For example, Hofstadter's butterfly, an intriguing pattern of the energy states of Bloch electrons, was predicted several decades ago to be observable only under unfeasibly strong magnetic fields in conventional materials. But it has been observed recently under current experimental conditions in graphene/BN layered heterostructures, one of the outstanding new kinds of 2D materials.[5-8] Moreover, another amazing physics phenomenon, Majorana fermions was predicted to exist in heterostructural systems consisting of a superconductor (SC) and a topological insulator (TI) Journal.[9-14] The coupling between an s-wave superconductor and the edge state of a 2D topological insulator can produce the long-sought Majorana quasiparticle excitations, with outstanding applications for high-efficiency quantum computation in the future.[9-22] The key to observing the Majorana phenomenon experimentally is finding ideal material systems, typically by construction of TI/SC hybrid structures. A few such heterostructures have been reported, fabricated by stacking different low-dimensional materials.[19-22] However, the transfer and stacking process involves complicated steps of advanced microfabrication techniques. Certainly, it is essential both to explore more related heterostructures and to develop new approaches, like a method based on spontaneous formation of desired heterostructures, for both fundamental research and technological applications.

In particular, we note that $HfTe_5$ films in few-layer thickness have recently been theoretically predicted as a promising large-gap topological insulator.[23] $HfTe_3$ bulk probably displays superconductivity.[24] Interesting, these two materials have a layered configuration in the *a-c* plane. In the *a-c* plane, a $HfTe_5$ layer can be considered as a $HfTe_3$ layer linked via Te-Te chains (see models in Fig. S1 in the Supplementary Information, SI).[23,25-27] These data from literatures provide hint for the possibility of the HfTe$_3$/HfTe$_5$ layered heterostructures with combined properties of topological insulator and superconductor. Besides the bulk materials, however, either HfTe$_3$ or HfTe$_5$ films are not reported experimentally yet. This stimulates us to explore a spontaneous way to construct such layered materials and their heterostructures. This idea may lay the groundwork for the construction of novel 2D systems with integrated properties in few-layer thickness, suitable for superconducting devices and experimental realization of Majorana fermions.

Here we report fabricating a superconductor- topological insulator-normal metal heterostructure with a layered configuration of HfTe$_3$/HfTe$_5$/Hf for the first time. By optimizing the experimental process, we find that this heterostructure can indeed form spontaneously. The atomic structure of the heterostructure has been determined by in-situ scanning tunneling microscopy (STM) and X-ray photoelectron spectroscopy (XPS). Scanning tunneling spectroscopy (STS) measurements directly reveal a bandgap as large as 60 meV in HfTe$_5$ film and a superconducting spectrum in HfTe$_3$/HfTe$_5$ film. And the combination of topological insulator and superconductor indicates that such a heterostructure can serve as an important platform for realizing many amazing phenomena.[9-14] Unlike the common film lift-transfer-stacking technique, our current method of making desired heterostructures is based on a spontaneous formation process of surface reaction and epitaxial growth and significantly simplifies the fabrication process. This method may offer new routes for the development of other related functional heterostructures and nanodevices.

The HfTe$_3$/HfTe$_5$ heterostructures were fabricated through direct reaction and epitaxial growth of tellurium atoms on a Hf(0001) substrate (see the schematic in Fig. 1). Our method for forming this heterostructure is a straightforward and transfer-free technique. Only tellurium atoms were deposited on a Hf(0001) substrate (Fig. 1a). The substrate was annealed to 500 ℃ while depositing Te atoms to obtain epitaxial $HfTe_5$ film (Fig. 1b). Then the substrate was annealed to 530~590 ℃ without depositing Te atoms to obtain epitaxial $HfTe_3$ film at the top of the sample (Fig. 1c).

To investigate the atomic structure of the as-grown epitaxial films described in Fig. 1b, we performed STM studies. Figure 2a is a large-scale STM image of the sample. The zoom-in image (Fig. 2b) shows a well-ordered pattern of protrusions with rectangular symmetry and crystalline constant of the *a-c* plane in $HfTe_5$ bulk material (more discussions are presented in Fig. S1 in SI).[23,25,26] Moreover, the formation of $HfTe_5$ film was monitored by in-situ X-ray photoemission spectroscopy (XPS). The XPS data from this $HfTe_5$ film were consistent with those of $HfTe_5$ single crystal (see Fig. S2 in SI). With a combination of these data, we conclude that a $HfTe_5$ film is formed on the substrate at this stage.

In order to further reveal the electronic properties of the $HfTe_5$ film, we performed STS measurements (Fig. 2c). STS detects the differential tunneling conductance (dI/dV), which gives a measure of the local density of states of the samples near the Fermi level (zero bias) of electrons at energy eV. The dI/dV spectra in Fig. 2c show that the Fermi level is within the energy gap. This energy gap is as large as 60 meV, and is close to the energy gap theoretically calculated for $HfTe_5$ with a thickness of three layers (see Fig. S3 in SI),[23] providing further clear evidence of the formation of $HfTe_5$ film.

We then conducted further thermal treatment of the sample with increasing sample temperatures to drive the formation of $HfTe_3$ film at the top of the substrate, as described in Fig. 1c. Figure 2d is a large-scale STM image of the treated sample. We can see that its topography is clearly different from that of $HfTe_5$ film shown in Fig. 2a. Figure 2e is a zoom-in image of the annealed sample. The well-ordered protrusions in this image exhibit a rectangular lattice (the model is shown in Fig. S1). The periodicities of these protrusions match the crystalline constants in the a-c plane in HfTe$_3$ bulk.[28] The electronic properties of HfTe$_3$ film were also investigated by STS at sample temperature of 4 K, as shown by the *dI/dV* spectrum in Fig. 2f, the feature in which is obviously different from that of HfTe$_5$ film (Fig. 2c). One significant difference is that no bandgap exists in the HfTe$_3$ film, while HfTe5 film has an energy bandgap of 60 meV.

To confirm and analyze the possible superconducting properties of the resulting HfTe$_3$ film, *dI/dV* spectra were further obtained at various sample temperatures and externally applied magnetic fields, which are often utilized together to determine a material's superconductivity. In the as-prepared HfTe$_3$ film, we indeed observed superconducting gap-like spectra: a pronounced dip in the density of states at the Fermi level, and one peak existing on each side of this dip (see *dI/dV* spectra in Figs. 3a, 3c and S4). Figure 3a shows the spectra measured at different sample temperatures from 0.48 to 0.93 K with interval of 0.1 K. From these experimental spectra, the superconducting gap ($\Delta$), critical temperature ($T_C$) for superconductor and the Bardeen-Cooper- Schrieffer (BCS) ratio $2\Delta/k_B T_C$ ($k_B$ is the Boltzmann constant) can be deduced by the BCS fitting,[29] the values are 0.20 meV, 1.23 K and 3.75, respectively (Fig. 3b). The superconducting property of HfTe$_3$ film is further supported by STS experiments under variable external magnetic fields. Figure 3c shows the dI/dV spectra obtained at magnetic fields from 0 to 6 T, applied to the sample surface in vertical direction. By a systematic analysis of the zero-bias conductance (ZBC) in these field-dependent dI/dV spectra, an upper critical field ($H_{C2}$) of 5.76 T for completely suppressing superconductivity can be obtained (Fig. 3d). Thus, these combined measurements verified the superconducting properties of the HfTe$_3$ film.

To gain further insight into the epitaxial film and its growth process, we performed in-situ XPS studies of the formation process of the heterostructure. Figure 4a shows the XPS spectra of the $Te_{3d}$ core level during the growth of the film with increasing sample temperatures. After the substrate was annealed to ~500 ℃ while depositing Te atoms, two peaks of binding energy (marked by the red at binding energies 582.82 and 572.42 eV) are the same as those of $HfTe_5$ single crystal (Fig. S5 in SI). These two peaks can be assigned to the $HfTe_5$ film, as revealed by the STM measurements at this stage (Fig. 2a). Further increasing the sample temperature up to 530 ℃ without depositing Te atoms, as described in Fig. 1c, results in decreased intensity of the red peaks (at 582.82 and 572.42 eV), while two new peaks appear (marked by the purple at binding energies 583.47 and 573.0 eV). These two purple peaks can be assigned to the $HfTe_3$ film, as analyzed by the STM measurements (Fig. 2d). Even at ~590 ℃, the intensity of purple peaks increases while the red peaks still exist, demonstrating the coexistence of $HfTe_5$ and $HfTe_3$ films.

Another issue is the stacking order of the $HfTe_3$ and $HfTe_5$ films. First, we found that the geometric structure of the whole sample surface is the same as the STM observation in Fig. 2d, indicating that the topmost layer of the sample is $HfTe_3$ film. Second, the observations of superconductivity of $HfTe_3$ film imply the existence of a buffer layer between the $HfTe_3$ film and the normal metal substrate, screening out the influence of the substrate's electrons upon superconductivity. This buffer layer must be $HfTe_5$ film, which already exists on the metal before the formation of $HfTe_3$ film. Furthermore, to clarify the coexistence and layer order of the $HfTe_3/HfTe_5$ films, the ratios of photoemission peak area of $Te_{3d}$ in $HfTe_5$ to $Te_{3d}$ in $HfTe_3$ are measured as a function of annealing time (Fig. 4b and Fig. S5). As shown by curves in the inset of Fig. 4b, the peak area of $Te_{3d}$ in $HfTe_3$ increases, while that in $HfTe_5$ decreases during the initial annealing stage. After an annealing time up to ~ 3 minutes, both the peak areas of Te in $HfTe_3$ and in $HfTe_5$ remain constant. We therefore conclude that there is a $HfTe_5$ film between $HfTe_3$ film and the substrate. So a combined analysis of XPS, STM and STS data demonstrates that a $HfTe_3/HfTe_5$ layered films is formed on the Hf substrate.

In summary, we successfully fabricated a heretofore-unexplored superconductor-topological insulator-normal metal heterostructure with a $HfTe_3/HfTe_5/Hf$ layered configuration through direct reaction and epitaxial growth of Te atoms on a Hf(0001) substrate. The geometric structures and electrical properties of the heterostructure were experimentally determined by XPS, STM and STS. This layered heterostructure has potential for studies of both the QSH effect and topological phase transitions.[23] Our newly demonstrated method opens up a route to fabricate heterostructures and nanodevices with a combination of multi-properties, particularly to form systems to study Majorana quasiparticle excitations and topological quantum computation.[30,31]

*Experiment section*

Sample preparation. The superconductor-topological insulator-normal metal $HfTe_3/HfTe_5/Hf$ heterostructures were fabricated in an ultrahigh vacuum chamber, with a base pressure of 2 x $10^{-10}$ mbar, equipped with standard molecular beam epitaxy (MBE) capabilities. The Hf(0001) substrate was cleaned by several cycles of Ar+ ion sputtering, followed by annealing until clean surface terraces were obtained in the STM images. Plenty of Te (Sigma, 99.999%) evaporated from a Knudsen cell was deposited onto the clean Hf(0001) surface at room temperature. The sample was subsequently annealed at proper temperatures to achieve different structures (at 500 ℃ for the $HfTe_5$ structure, and at about 560 ℃ for $HfTe_3/HfTe_5$ layered structures). After growth, the samples were transferred to STM equipment and cooled for imaging of topographies and measuring of the local electronic properties.

XPS Measurements. The in-situ X-ray photoelectron spectroscopy measurements were performed at the Beijing Synchrotron Radiation Facility (BSRF). Synchrotron radiation light monochromated by four high-resolution gratings and controlled by a hemispherical energy analyzer has photon energy in the range from 10 eV to 1100 eV.

STM Measurements. To obtain the electronic structures of the films, STM experiments were performed using an ultralow-temperature STM system (Unisoku and RHK) operated at a base temperature of 0.45 K and under a magnetic field up to 9 T. The differential conductance ($dI/dV$) was measured using lock-in detection of the tunnel current I by adding a 5 mV for 4 K and 0.15 mV for 0.45 K modulated bias voltage at 973 Hz to the sample bias voltage V. The energy resolution is better than 0.4 meV at 0.45 K with an electrochemically etched tungsten tip.

Acknowledgements

We acknowledge financial support from the National Basic Research Program of China (No. 2013CBA01600), National Natural Science Foundation of China (Nos. 61222112, 51572290, 51325204, and 11334006), and Chinese Academy of Sciences (Nos. 1731300500015 and XDB07030100).

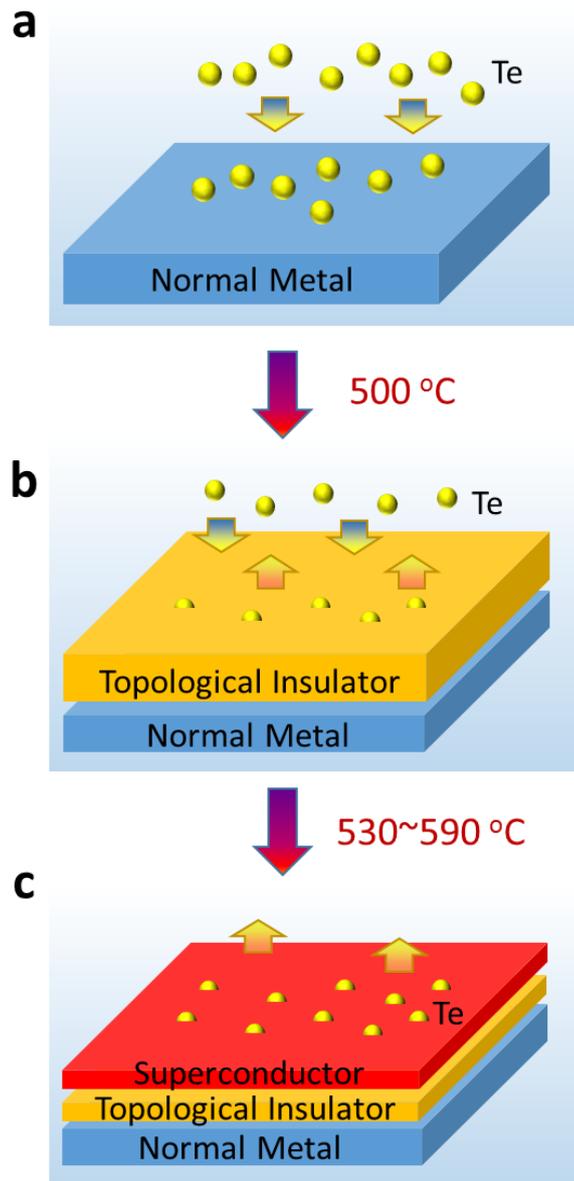

**Figure 1.** Schematic of the fabrication process. A superconductor- topological insulator- normal metal heterostructure is constructed by a single step of direct reaction and epitaxial growth of tellurium atoms on a Hf(0001) substrate. (a) Tellurium atoms are deposited on Hf(0001) substrate at room temperature. (b) The substrate is then annealed to 500 ℃ during deposition of Te atoms, and epitaxial insulated $HfTe_5$ film (indicated in yellow) is obtained. (c) While the substrate is further annealed to 530~590 ℃ without deposition of Te atoms, the topmost layer of the $HfTe_5$ film transforms into superconductive $HfTe_3$ film (indicated in red)

due to some of the Te atoms escaping from the $HfTe_5$ film. As a result, a $HfTe_3$/$HfTe_5$/Hf layered heterostructure is obtained.

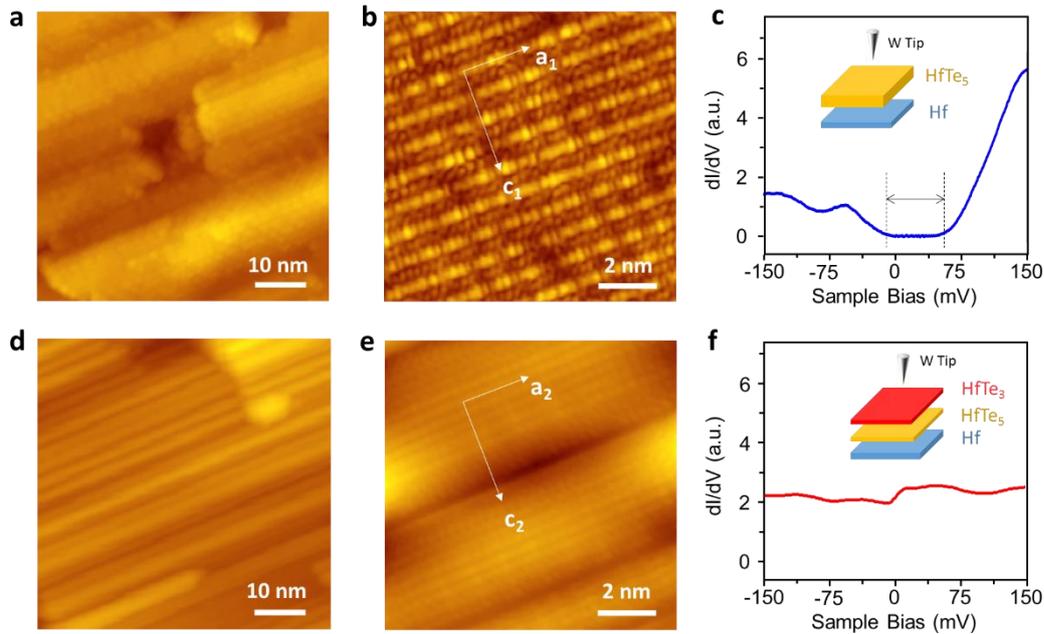

**Figure 2.** Structural and electronic properties of the as-grown films, obtained by STM/STS. (a) STM topographic image (-1.00 V, -1.00 pA) of the $HfTe_5$ film. (b) High-resolution image (50.0 mV, 1.00 nA) of the $HfTe_5$ film. (c) *dI/dV* spectrum (setpoint, -200 mV, -100 pA) taken at $HfTe_5$ surface. The black arrow and the dashed lines indicate the bottom of the conduction band (right) and the top of the valence band (left), respectively. (d) Topographic image (-5.00 V, -10.0 pA) of the $HfTe_3$ film. (e) High-resolution image (-200 mV, -500 pA) of the $HfTe_3$ film. (f) *dI/dV* spectrum (setpoint, -300 mV, -100 pA) taken at $HfTe_3$ surface. The arrows in b and e indicate the close-packed symmetric crystalline directions in the *a-c* plane. Insets in c and f show schematics of a tungsten tip located at the surface of the $HfTe_5$/Hf and a $HfTe_3$/$HfTe_5$/Hf heterostructure, respectively.

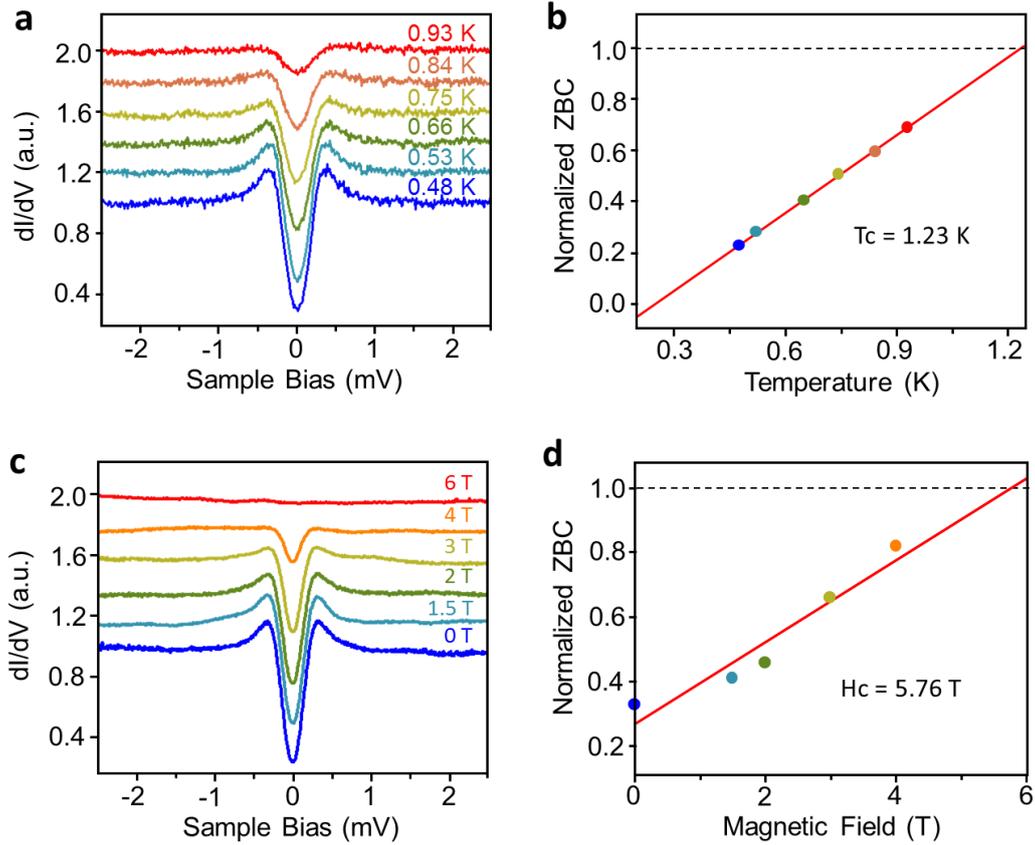

**Figure 3.** Superconductivity on the as-grown HfTe$_3$ film. (a) Differential conductance measured with a W tip on the HfTe$_3$ film as a function of sample temperatures. The curves at different temperatures are offset vertically for clarity. (b) The open circles show the measured gap and the solid curve shows the fitting by the BCS gap function. (c) The differential conductance measured on the HfTe$_3$ film as a function of external magnetic field. The tunneling junction (also applies to a) was set at $V$=40 mV and $I$=200 pA. (d) The dots show the normalized ZBC from the spectra in c. The upper critical field of the HfTe$_3$ film for completely suppressing superconductivity is estimated to be 5.76 T.

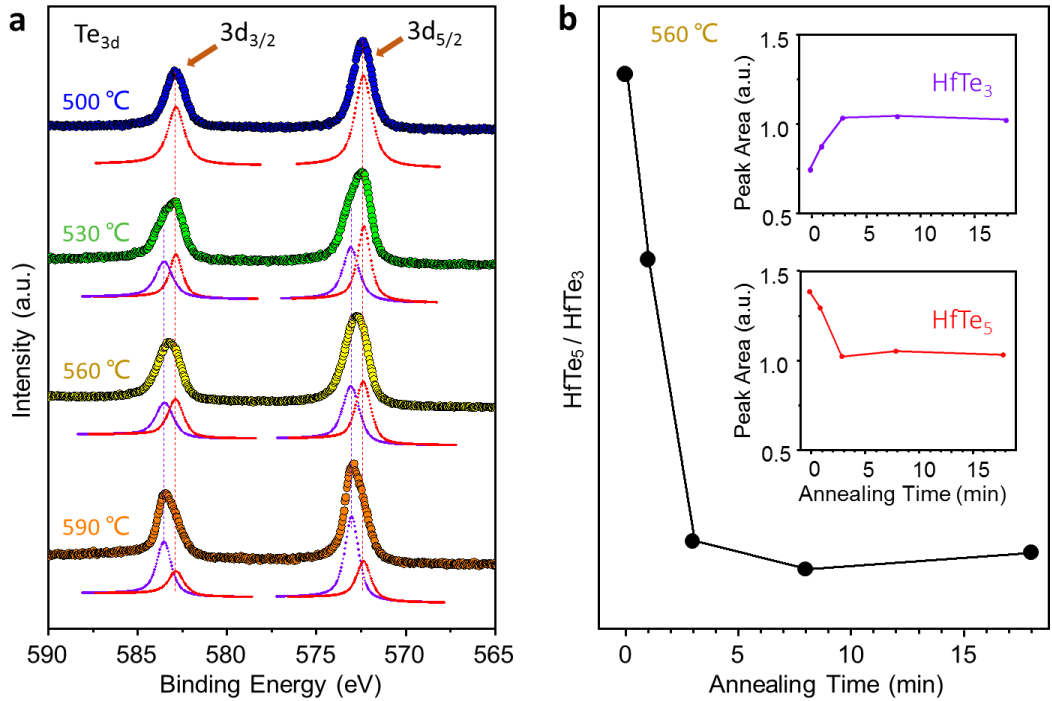

**Figure 4.** XPS measurements during the evolution of the heterostructure. (a) The binding energies of Te$_{3d}$ as a function of the sample temperature. The red peak positions (582.82 and 572.42 eV) correspond to the binding energy of Te$_{3d}$ in HfTe$_5$. The purple peak positions (583.47 and 573.0 eV) correspond to the binding energy of Te$_{3d}$ in HfTe$_3$. (b) Area ratio of Te$_{3d}$ in HfTe$_5$ to Te$_{3d}$ in HfTe$_3$ as a function of annealing time (the sample temperature was kept at 560 ℃). Insets show the normalized peak area of Te$_{3d}$ in HfTe$_5$ (red dots and line) and HfTe$_3$ (purple dots and line), respectively.

TOC

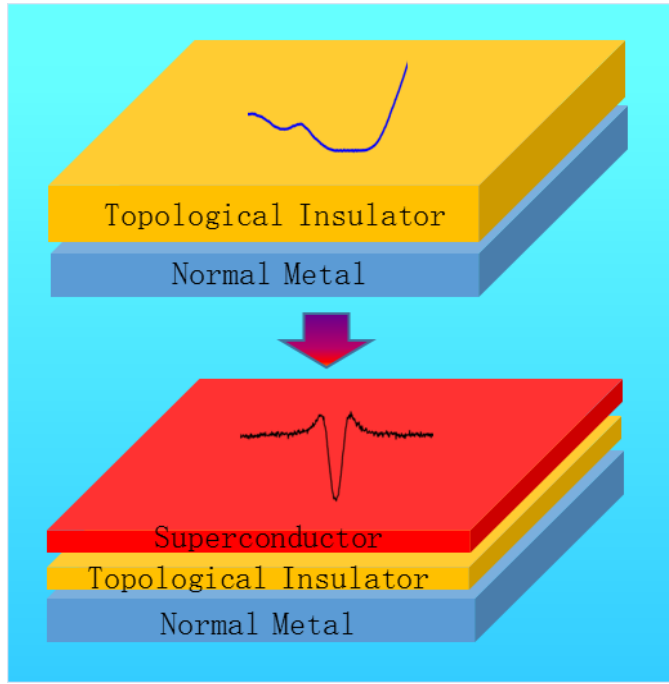

Two-dimensional (2D) materials with heterolayered structures beyond graphene are being explored due to the anticipation of unprecedented properties or multiple functions. For instance, heterostructures for possible realization of quantum spin Hall effect (QSHE) and Majorana fermions are urgently sought. Here a superconductor-topological insulator-normal metal heterolayered structure was reported. A combination of experimental characterizations reveals this hybrid structure to be a $HfTe_3/HfTe_5/Hf$ layered configuration, with potential applications in both QSHE and Majorana-based devices. In contrast to the familiar lift-transfer-stacking fabrication method of heterostructures, the current method is based on the spontaneous formation process: there is no lift-off task, and only Te atoms are deposited onto a hafnium substrate. This method gives inspiration to form desired heterostructures and nanodevices for novel technological applications like high-efficiency quantum computation.

# Supporting Information

## Spontaneous Formation of a Superconductor-Topological Insulator-Normal Metal Layered Heterostructure

By *Yu-Qi Wang, Xu Wu, Ye-Liang Wang\*, Yan Shao, Tao Lei, Jia-Ou Wang, Shi-Yu Zhu, Haiming Guo, Ling-Xiao Zhao, Gen-Fu Chen, Simin Nie, Hong-Ming Weng, Kurash Ibrahim, Xi Dai, Zhong Fang, Hong-Jun Gao\**

Beijing National Laboratory of Condensed Matter Physics, Institute of Physics, Chinese Academy of Sciences, Beijing 100190, P. R. China

Institute of High Energy Physics, Chinese Academy of Sciences, Beijing 100049, P. R. China.

Collaborative Innovation Center of Quantum Matter, Beijing 100084, P. R. China.

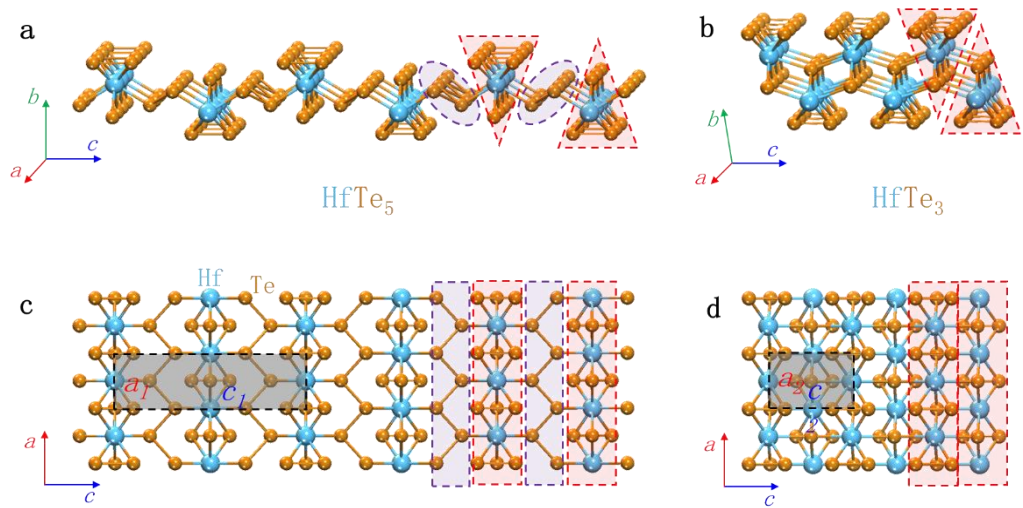

**Figure S1.** (**a**) and (**b**) 3D view of $HfTe_5$ and $HfTe_3$ layers, respectively. (**c**) and (**d**) The top view (*a-c* plane) of $HfTe_5$ and $HfTe_3$ layers, respectively. The crystalline constant of the *a-c*

plane is $a_1$=3.96 Å, $c_1$=13.68 Å in HfTe$_5$ layers and $a_2$=3.90 Å, $c_2$=5.87 Å in HfTe$_3$ layers.[S1-3] The HfTe$_5$ layer can be described as trigonal prismatic chains of HfTe$_3$ [marked by red dashed-line triangles in 3D view in **a, b** and by dashed-line rectangles in top view in **c, d**] linked via zigzag Te-Te chains [marked by purple dashed-line ovals in **a** and rectangles in **c**]. The unit cells in the *a-c* plane are marked by black dashed-line rectangles in **c** and **d**, corresponding to the STM observations in Figs. 2b and 2e in the main text, respectively.

The bond energy of the Te-Te chains in HfTe$_5$ is much weaker than that of the trigonal prismatic chains. So the Te-Te chains decompose much more easily than the Te-Hf keys. After the Te-Te chains decompose and the trigonal prismatic chains connect together one-by-one along the *c* direction, HfTe$_5$ layers turn into HfTe$_3$ layers (See the schematics from Fig. S1a to Fig. S1b). So that is probably the main reason we see fold-like corrugations in the HfTe$_3$ film, as revealed by STM measurements in Figs. 2d and 2e in the main text. The periodicities of the STM-observed protrusions of HfTe$_3$ film are 3.74 and 4.65 Å in the *a* and *c* directions, respectively (Fig. 2e). And the reported crystal lattice parameters are 3.90 and 5.87 Å in HfTe$_3$ bulk, respectively. The value in the *a* direction, determined from our STM images, almost matches that of the bulk, and the difference is probably due to the difference of the sample temperatures: STM images are taken at liquid helium temperature of about 4 K, while the reported bulk parameters are taken at room temperature. From the STM images, we can see that the corrugation in the *c* direction in the HfTe$_3$ film is larger, which probably means that the measured periodicity of the protrusions in the *c* direction is larger than that of the bulk.

Differences also exist between our STM observations of crystal lattice parameters in HfTe$_5$ film at liquid helium temperature and those reported for HfTe$_5$ bulk at room temperature. The parameters observed in HfTe$_5$ film are smaller than those in HfTe$_5$ bulk, similar to the case of HfTe$_3$. In the HfTe$_5$ film, the average periodicity of the protrusions in the *a* direction (Fig. 2a) is about 3.4 Å at liquid helium temperature, which is different from the reported periodicity (3.96 Å) in the bulk material at room temperature. The average periodicity of the protrusions in the *c* direction is about 10.9 Å at liquid helium temperature, which is also smaller than the average distance (13.68 Å) reported for the bulk material at room temperature.

The difference in crystal lattice parameters between the measured values in $HfTe_5$ film and $HfTe_5$ bulk is larger than that between $HfTe_3$ film and $HfTe_3$ bulk. The reason is probably the 'soft' Te-Te chains in $HfTe_5$, in which the bond energy is much weaker than that of the trigonal prismatic chains. Thus the zigzag Te-Te chains in $HfTe_5$ that link trigonal prismatic chains easily become deformed at liquid helium sample temperature, constraining the $HfTe_5$ film. The crystal lattice parameters, especially in the *c* direction, accordingly become smaller than those of $HfTe_5$ bulk at room temperature. In addition, the constrained of the lattice probably leads to the change of the surface electronic state in the film. Therefore, the changes of the geometric and electronic structures are the main reasons the protrusions' periodicity observed in STM image differs from the lattice parameters in bulk.

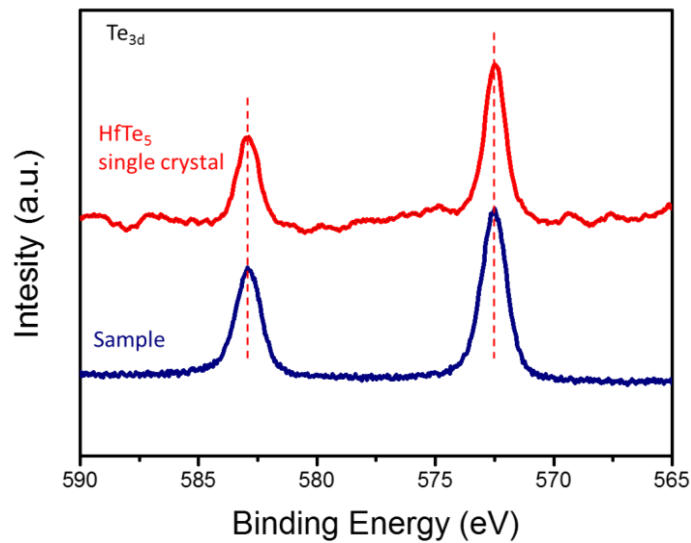

**Figure S2**. XPS measurements for the binding energies of Te$_{3d}$ in the HfTe$_5$ film while annealing sample to ~500 ℃ (blue curve) and in a HfTe$_5$ single crystal (red curve). In these two spectra the peaks' positions (582.82 and 572.42 eV) match well, as marked by red dashed-line. This agreement proves the existence of HfTe$_5$ film at surface of the sample annealed to 500 ℃, as revealed by STM and STS data (Figs. 2a-c) in the main text.

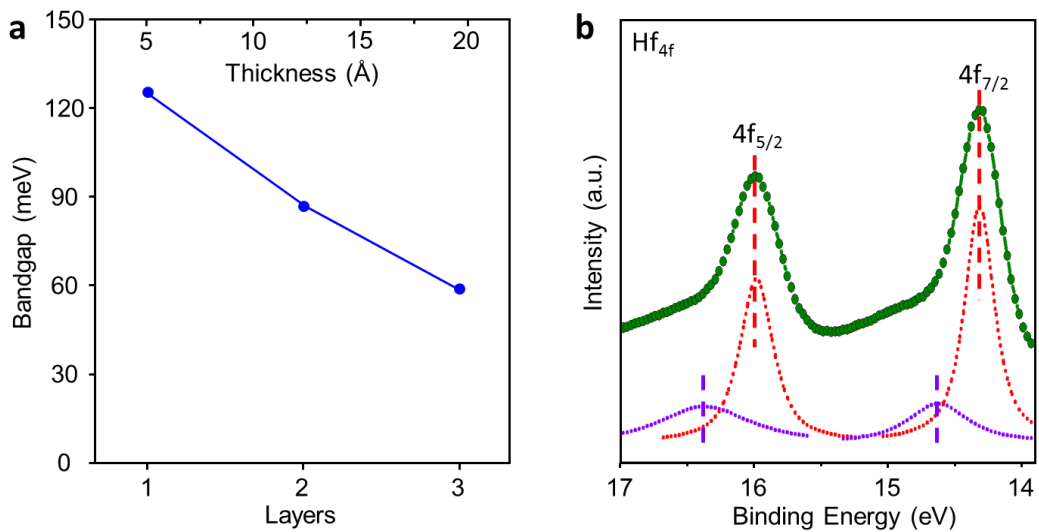

**Figure S3**. (**a**) Calculated bandgap of HfTe$_5$ as a function of the number of layers (the computational methods are described in Ref.S1). The bandgap of HfTe$_5$ decreases with the increasing number of layers. The calculated thickness of each layer is 0.72 nm. (**b**) XPS measurements for the binding energies of Hf$_{4f}$ at sample temperature of 500 ℃. The red peak positions (15.96 and 14.29 eV) correspond to the binding energy of Hf$_{4f}$ in Hf substrate. The purple peak positions (16.35 and 14.60 eV) correspond to the binding energy of Hf$_{4f}$ in HfTe$_5$. From these XPS data, we can get the area ratio of Hf$_{4f}$ in HfTe$_5$ to Hf$_{4f}$ in Hf substrate is 0.7. The escape depths in HfTe$_5$ and Hf$_{4f}$ in Hf are calculated as about 1.3 nm by the NIST program IMFPWIN. Based on these data, we obtained the thickness of our HfTe$_5$ film with a value of 2.1 nm. The bandgap from this film which is measured by *dI/dV* spectrum as large as 60 meV (Fig. 2c), which is almost the same as the bandgap theoretically calculated for HfTe$_5$ film with three layers (2.16 nm in thickness).

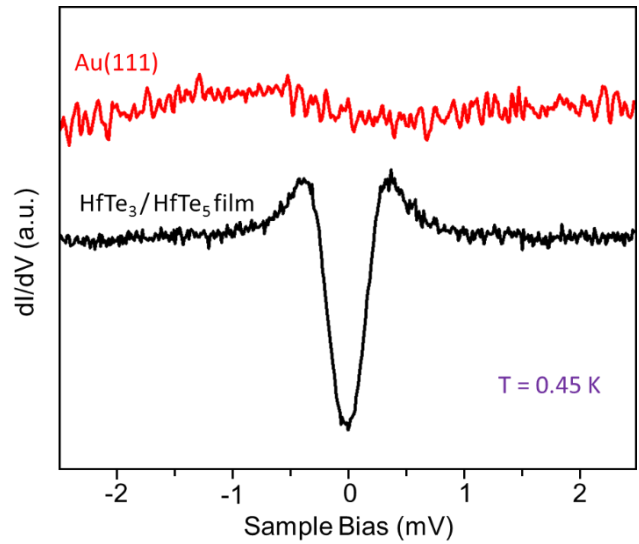

**Figure S4.** Differential tunneling conductance (*dI/dV*) measured at the HfTe$_3$/HfTe$_5$ film (black curve) and at the Au(111) surface (red curve) with a tungsten tip at temperature of 0.45 K. The *dI/dV* data obtained at the Au(111) surface show standard gold spectra,[S4] in which a clear feature is the absence of peaks and dips at the Fermi level. So the black superconducting gap-like spectrum observed in HfTe$_3$/HfTe$_5$ film originates from the HfTe$_3$/HfTe$_5$ film itself, and has nothing to do with the tip's electronic state.

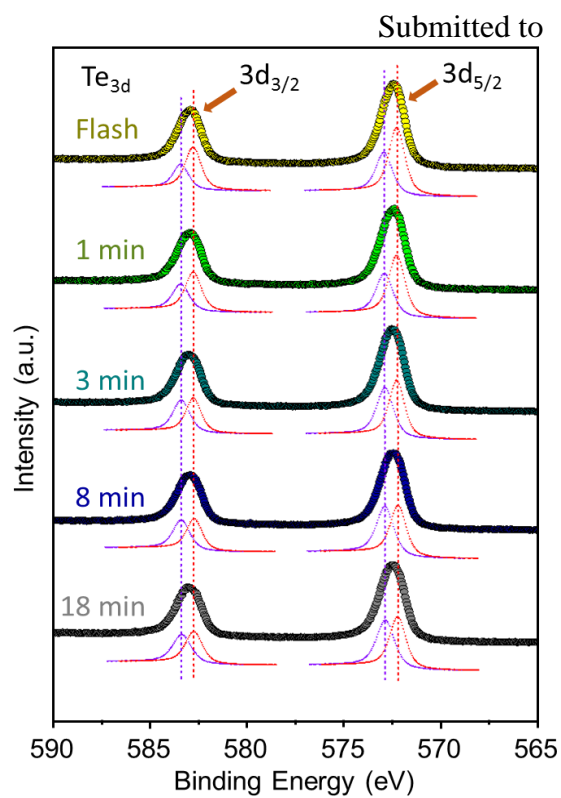

**Figure S5**. XPS measurements for the binding energies of Te$_{3d}$ at sample temperature of 560 ℃. Cycles and related curves are raw experimental data with increasing time, which can be fitted into the red and purple peaks in each curve, as demonstrated in Fig. 4a. The red ones indicate the peak positions (582.82 and 572.42 eV) corresponding to the binding energy of Te$_{3d}$ in HfTe$_5$. The purple ones indicate the peak positions (583.47 and 573.0 eV) corresponding to those in HfTe$_3$. We can clearly see that the peak areas of Te$_{3d}$ remain constant after ~3 minutes. The peak area ratios of red curve to purple curve as a function of time are presented in Fig. 4b.